%% file: paper.tex
\documentclass{vgtc}                          

\ifpdf
  \pdfoutput=1\relax                   
  \usepackage{graphicx}                
  \DeclareGraphicsExtensions{.pdf,.png,.jpg,.jpeg} 
\else
  \usepackage{graphicx}                
  \DeclareGraphicsExtensions{.eps}     
\fi%

\graphicspath{{figures/}{pictures/}{images/}{./}} 

\usepackage{microtype}                 
\PassOptionsToPackage{warn}{textcomp}  
\usepackage{textcomp}                  
\usepackage{mathptmx}                  
\usepackage{times}                     
\usepackage{cite}                      
\usepackage{tabu}                      
\usepackage{booktabs}                  

\onlineid{1047}

\vgtccategory{Research}

\vgtcinsertpkg

\input{custom}

\title{Dataspace: A Reconfigurable Hybrid Reality Environment \\ for Collaborative Information Analysis}

\author{Marco Cavallo\thanks{e-mail: marco@mastercava.com}\\ %
        \scriptsize IBM Research %
\and Mishal Dholakia\thanks{e-mail: mishal.dholakia1@ibm.com}\\ %
     \scriptsize IBM Research %
\and Matous Havlena\thanks{e-mail: havlenam@ibm.com}\\ %
     \scriptsize IBM Research %
\and Kenneth Ocheltree\thanks{e-mail: kochel@us.ibm.com}\\ %
     \scriptsize IBM Research %
\and Mark Podlaseck\thanks{e-mail: podlasec@us.ibm.com}\\ %
     \parbox{1.4in}{\scriptsize \centering IBM Research}}

\abstract{
\input{abstract}
} 

\CCScatlist{
  \CCScatTwelve{Human-centered computing}{Visualization}{Visualization Systems and Tools}{};
  \CCScatTwelve{Human-centered computing}{Human computer interaction (HCI)}{Interactive systems and tools}{}
}

\begin{document}


\firstsection{Introduction}

\maketitle


\input{introduction}

\input{related}

\input{design}

\input{method}

\input{applications}

\input{discussion}

\input{conclusion}

\bibliographystyle{abbrv-doi}

\balance

\bibliography{paper}
\end{document}

%% file: custom.tex
\usepackage{enumitem}
\usepackage{balance}
\usepackage{cancel}
\DeclareRobustCommand{\bfparhead}[1]{\noindent\textbf{#1}} 

\usepackage{enumitem}
\setlist[itemize]{leftmargin=4mm}

%% file: abstract.tex
Immersive environments have gradually become standard for visualizing and analyzing large or complex datasets that would otherwise be cumbersome, if not impossible, to explore through smaller scale computing devices. However, this type of workspace often proves to possess limitations in terms of interaction, flexibility, cost and scalability.
In this paper we introduce a novel immersive environment called \textit{Dataspace}, which features a new combination of heterogeneous technologies and methods of interaction towards creating a better team workspace.
Dataspace provides 15 high-resolution displays that can be dynamically reconfigured in space through robotic arms, a central table where information can be projected, and a unique integration with augmented reality (AR) and virtual reality (VR) headsets and other mobile devices.
In particular, we contribute novel interaction methodologies to couple the physical environment with AR and VR technologies, enabling visualization of complex types of data and mitigating the scalability issues of existing immersive environments.
We demonstrate through four use cases how this environment can be effectively used across different domains and reconfigured based on user requirements.
Finally, we compare Dataspace with existing technologies, summarizing the trade-offs that should be considered when attempting to build better collaborative workspaces for the future.

%% file: introduction.tex
\begin{figure*}
  \centering
  \includegraphics[width=2\columnwidth]{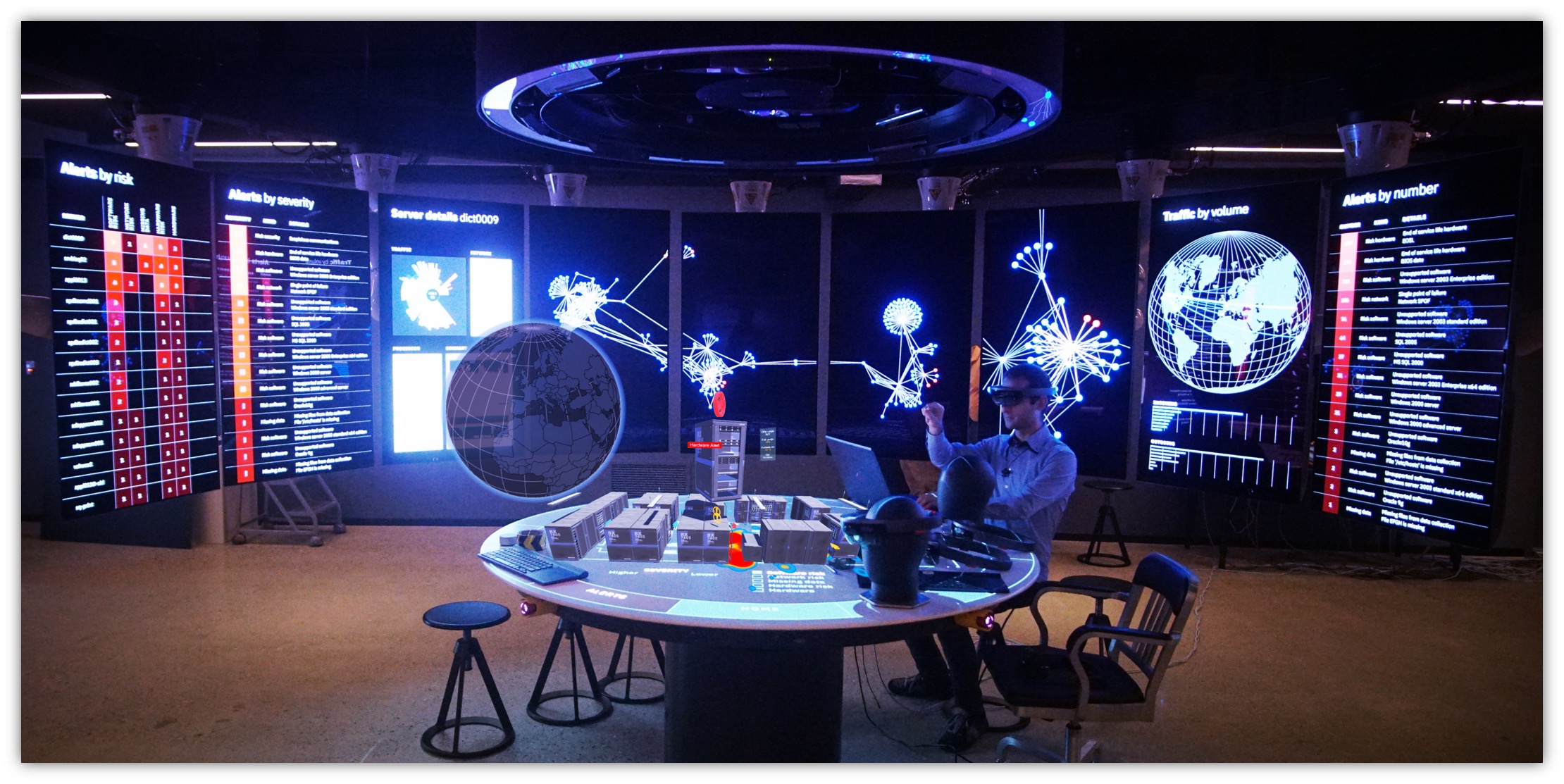}
  \caption{Image of the DataSpace. Dataspace is an immersive, collaborative, and reconfigurable environment, combining heterogeneous technologies and mixed interaction methodologies. The workspace includes 15 large high-resolution displays attached to moving robotic arms, two table projectors, and is complemented by the integration of augmented reality (AR) and virtual reality (VR) headsets, laptops and other mobile devices.
The figure above shows a sample use of the environment for our Data Center Management use case. In this particular configuration, the screens are positioned vertically, and display information on system faults and performance in a data center. A projection on the table displays a circular user interface that the user may touch to perform different selections on the data; tridimensional and/or aggregate information (floor plan, physical location of servers, etc.) is made available through an augmented reality headset. Please note that the AR pictures presented in this paper were captured with a DSLR camera through the Hololens Spectator View \cite{spectatorview} method.}~\label{fig:datacenter}
\vspace{-2em}
\end{figure*}

In recent years, the scientific community has faced an exponential increase in the amount of data being digitally collected and stored. 
This massive shift has challenged researchers to find ways to make sense of these vast datasets, rendering the development of visualization instruments increasingly essential.
Summarization, abstraction, and focus-plus-context techniques can be used to 
provide overview information on data and its structure, resulting in attempts to fit vast quantities of data on a single display. However, limited screen size and resolution have been demonstrated to be detrimental to multi-scale exploration of large datasets \cite{ni2006increased,reda2015effects}, while the concept of surrounding the user with visuals (``immersion'') has proven to be effective in the analysis of spatial data \cite{mcintire2014possible}.
Based on these considerations, high-resolution tiled display walls (also known as ``powerwalls'') \cite{renambot2004sage,marrinan2014sage2} and immersive virtual environments \cite{cruz1993surround} were introduced for exploring 2D and 3D datasets, respectively, between the '90s and the early 2000s. More recent environments such as CAVE2 \cite{febretti2013cave2} and Reality Deck \cite{papadopoulos2015reality} have attempted to combine the advantages of both powerwalls and immersive environments by providing incremental upgrades, but much work still needs to be done to improve the flexibility, scalability and reproducibility of these workspaces, which are often complex to build, hard to maintain and often cost prohibitive. \\
We introduce \textit{Dataspace} as a novel solution to large-scale data visualization, re-imagining the conference room as a dynamic physical environment for experiencing complex data and jointly making better-informed decisions.
Dataspace is a room-sized collaboration environment where people can work together and interact naturally with both 2D and 3D information, leveraging a unique combination of movable high-resolution displays, an interactive projection table, and augmented and virtual reality head-mounted displays. Unlike its predecessors, Dataspace physically adapts to various user, application, and data contexts, and is specifically aimed at improving the collaborative decision-making process. Additionally, the design of this environment attempts to mitigate the usual drawbacks of immersive virtual environments, especially with respect to reproducibility and extensibility.
In this paper we specifically contribute:
\begin{itemize}[noitemsep,topsep=0pt]
\item A set of high-level guidelines for the creation of data-oriented collaborative workspaces
\item The design and implementation of our new reconfigurable hybrid reality environment, Dataspace
\item A novel approach to integrating augmented and virtual headsets into an immersive environment.
\end{itemize}

After describing our proposed system, we analyze its application to four different data analysis use cases and settings, emphasizing its added value to the collaborative generation of new insights.
Finally, we discuss how Dataspace compares to existing immersive environment solutions, evaluating their technical features and use case applications of each, and promoting shared, flexible data visualization as the founding basis for the meeting room of the future.

%% file: related.tex
\section{Related Work}

Dataspace draws on prior work on immersive environments and collaborative systems, especially in data analytics. It also incorporates interaction and data visualization principles specific to augmented and virtual reality.

\subsection{Immersive Environments and Tiled Display Walls}
Scientists have been building systems for exploring 3D spatial data such as molecules, astrophysical phenomena, and geoscience datasets since the early ‘60s. Starting with Sutherland's work \cite{sutherland1965ultimate} in 1965, research has focused primarily on the development of single-user, desk-based immersive visualization systems, generally consisting of bulky head mounted displays with limited resolution and field of view.
With data continuing to grow in complexity, visualization instruments have become increasingly essential in conducting meaningful research. Influenced by advances in human-computer interaction \cite{krueger1983artificial}, the introduction of CAVE (CAVE Automatic Virtual Environment) \cite{cruz1993surround} in 1992 provided new ways to help lead users from raw data to discovery. Consisting of a cube measuring 10 feet at each side, CAVE leveraged a set of projectors to allow a small number of researchers to experience stereo 3D graphics on five of its sides (floor included). Marker-based head tracking and the use of lighter stereo glasses enabled users to physically move around in the environment while exploring the data, and the introduction of a tracked 3D ``wand'' provided an initial foray into embodied interaction.
CAVE successfully fostered scientific discovery for years without compromising the color, resolution, and flicker-free qualities of existing single-screen stereo graphics workstations.\\
Thanks to improvements in LCD technologies, the early 2000s saw the rise of tiled display walls as a more practical platform for large-scale visualization. Often spanning large surfaces, LCD display walls offered superior image quality and resolution with relatively low maintenance. They also enabled the visualization of large datasets while providing both detail and context, and opened up new possibilities for collaborative data analysis, as demonstrated by the use of SAGE \cite{renambot2004sage} (later SAGE2 \cite{marrinan2014sage2}) in the EVL Cybercommons room \cite{marai2016interdisciplinary} and by various studies on shared interaction \cite{jakobsen2014up,liu2016shared,prouzeau2017evaluating}. Similarly, LambdaTable \cite{krumbholz2005lambda} provided a first attempt to merge table interaction and tangible user interfaces with high-resolution tiled display technology.\\
Almost 20 years after its predecessor debuted, CAVE2 \cite{febretti2013cave2} was introduced as a response to the exponential increase in data gathered by the scientific community, either from observation of natural phenomena or from complex supercomputer simulations. The new ``virtual reality theater'', composed of 72 cylindrically positioned displays, aimed at combining the effectiveness of CAVE systems in visualizing 3D datasets with the capabilities of more recent ultra-high-resolution environments, which were a better fit for 2D data visualization. By combining the SAGE \cite{renambot2004sage} tiled display system and of OmegaLib \cite{febretti2014omegalib} virtual reality middleware, CAVE2 enabled researchers to seamlessly interact with large collections of 2D and 3D data, providing the first full implementation of a Hybrid Reality Environment (HRE) \cite{febretti2013cave2}.
In recent years, researchers have explored a number of ways to further improve on the CAVE legacy \cite{defanti2011future}. Focusing on the visualization of ultra-high-resolution geospatial data, Reality Deck \cite{papadopoulos2015reality} managed to increase the environment's total resolution from 36 MegaPixels to more than 1 GigaPixel by tiling 416 LCD displays together. DSCVR \cite{ponto2015dscvr} focused instead on a smaller number of displays, attempting to tackle concerns surrounding the scalability and reconfigurability of the environment, while other researchers tried to minimize production costs without renouncing the original CAVE design \cite{lau2007virtual,defanti2009starcave,amatriain2009allosphere,kageyama2013applications}. Parallel efforts attempted to improve users' interactions with the data, sometimes integrating external devices such as tablets \cite{krum2014tablet} and smart watches \cite{horak2018david}. Finally, inspired by the rise of new Artificial Intelligence technologies, the Cognitive Environments Lab (CEL) \cite{farrell2016symbiotic, venkataraman2016ceding} explored new cognitive interaction methodologies for collaboration, remote participation, and agent-based data exploration. \\
Our environment, Dataspace, attempts to bring together the advantages of all these workspaces. Dataspace can be considered the second full implementation of a Hybrid Reality Environment after CAVE2 \cite{febretti2013cave2}, combining high-resolution displays with augmented reality headsets for visualizing 3D data, plus a central interactive table and the integration of AI-based cognitive functionalities. Additionally, Dataspace differentiates itself from its predecessors with a design aimed at spatial reconfigurability, scalability, extensibility, and egalitarian access to data.

\subsection{Towards HMD Based Immersive Analytics}
While CAVE was originally introduced as an alternative to head-mounted displays (HMDs), in recent years great improvements have been made to virtual and augmented reality headsets, which now provide decent resolution and field of view at affordable prices and in reasonable form factors. 
As outlined by Papadopoulos et al. \cite{papadopoulos2015scalability}, immersive environments like CAVE are characterized by high costs, complex maintenance and scalability limits --- problems that can now be addressed by HMD-based Immersive Analytics \cite{cliquet2017towards}.
HMDs have successfully been applied to the visualization of brain information \cite{he2017medical} and other scientific data, and have been used in a variety of data visualization settings, including smartphone-based VR \cite{butcher2016immersive}, desk VR applications \cite{zielasko2017buenosdias}, and marker-based AR \cite{ritsos2017synthetic}.
Millais et al. \cite{millais2018exploring} demonstrate the advantages of using immersion for data exploration in virtual reality, while Butscher et al. \cite{butscher2018clusters} examine how immersive technologies can facilitate collaborative analysis to better detect clusters, trends and outliers.
Cordeil et al. \cite{cordeil2017immersive} demonstrate that modern HMDs provide a comparable experience to CAVE-style facilities for collaborative abstract data analysis. 
However, as outlined by McIntire et al. \cite{mcintire2014possible}, the use of stereoscopic displays alone for information visualization still has its limitations. In particular, AR and VR headsets can be convenient for performing tasks associated with spatial or multidimensional data, but fall short in displaying statistical and abstract information, which is instead more successfully handled by 2D visualizations \cite{bach2016immersive}. \\
While both appreciating these technologies and also acknowledging their limitations, we have seamlessly integrated AR headsets into our Dataspace environment, enabling collaborative analysis of spatial datasets and allowing the users to rapidly move from high-resolution 2D information displayed on the Dataspace screens to 3D data visualized in AR (and vice-versa). Similarly, we employ VR headsets to create a ``virtual extension'' of the environment for remote participation.

%% file: design.tex
\section{Design Criteria}
Taking into account the literature, plus previous experiences working with environments such as CAVE2 \cite{febretti2013cave2} and CEL \cite{farrell2016symbiotic}, we devised a set of design criteria to apply to the creation of a data-oriented collaborative workspace. We refer back to these criteria throughout the paper while describing our Dataspace implementation.

\bfparhead{D1. Shared Data Exploration} The ability to solve complex problems involving big data requires a wide range of skills. Therefore, the environment should provide an inviting space where scientists of various backgrounds can comfortably sit together, analyze, and interpret data, facilitating synergistic insights
better than those a lone individual might arrive at. When possible, the space should encourage both co-located and remote scientists to collaborate.

\bfparhead{D2. Egalitarian Access to the Data}
While existing systems sometimes situate their data explorations in a particular viewpoint (CAVE2's 3D rendering, for example, is always based on the perspective of only one user at a time \cite{febretti2013cave2}), we provide egalitarian access to data through a full circular design and related software solutions. We encourage more people to become involved in the shared experience, independent of their physical position in the environment.
When required by the application, we of course also provide tools and interactions for individuals to assume various roles throughout data exploration (e.g. orienting or focusing content and lighting in a particular direction, responding to the current speaker).

\bfparhead{D3. Flexible Data Immersion}
Using a number of different applications results in a mixture of different types of visual content, each associated with unique interactions, focus, and level of data immersion. On one side of this spectrum we can imagine observing a 2D image on a large screen from the distance, while on the other end we can imagine being completely immersed in the data through a virtual reality headset while performing direct manipulation. Thanks to its ability to reconfigure the position of screens in space, and AR/VR integration, Dataspace provides extreme flexibility across the whole data immersion spectrum via both software and hardware solutions. This includes support for hybrid rendering of both 2D and 3D information simultaneously, a feature whose importance was already noted by Febretti et. al \cite{febretti2013cave2}.

\bfparhead{D4. Multimodal Interaction}
The system should support a variety of naturalistic interactions, such as keyboard/mouse, touch, spatial controllers (6 DOF joysticks, wands), and voice-activated interfaces. While each application may make a different use of these interactions, we believe in the importance of making them all available to the user, so that he can interact through touch, voice, or other means, depending on application context, personal preference, spatial position, and environment-specific properties (e.g. team members, background noise). This is also important because it provides a number of fallbacks in case of a system failure.

\bfparhead{D5. Seamless Integration of Heterogeneous Devices}
A fundamental aspect to consider while designing a multi-purpose collaborative system is the integration with external devices. While touch or voice interactions could be sufficient for certain applications, other tasks may me be performed more efficiently through the adoption of other application-specific devices, either standard (e.g. a physical keyboard for typing) or less conventional (e.g. augmented or virtual reality headsets for 3D data visualization and spatial interactions). It is important to generalize system interactions and content delivery in order to simplify extension to new devices and uses.

%% file: method.tex
\section{DataSpace}


\begin{figure}
  \centering
  \includegraphics[width=\columnwidth]{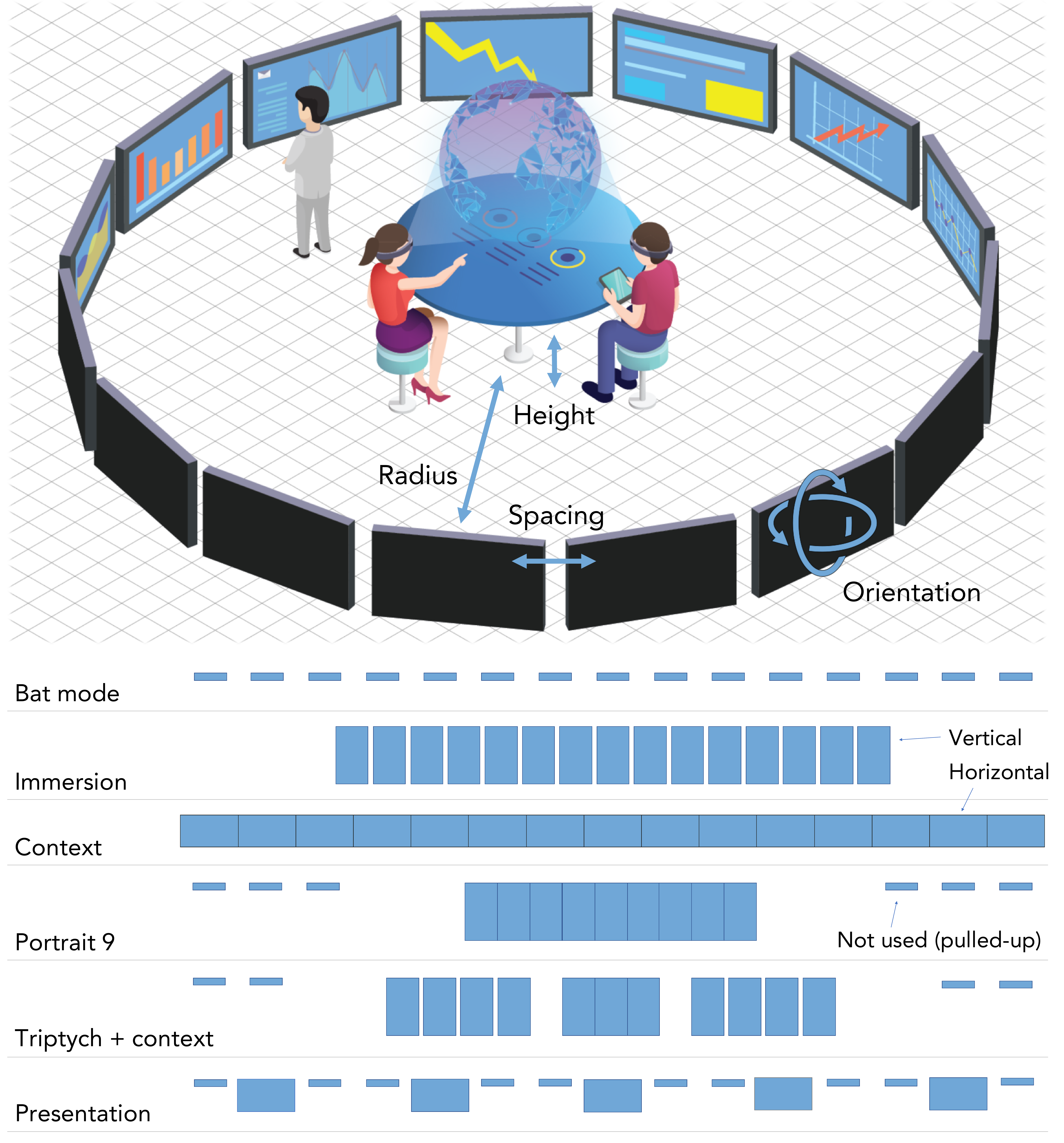}
  \caption{Sample screen configurations. Thanks to the 7-DOF robotic arms, screens can be dynamically positioned and oriented in space. Typical configurations include screens in portrait (e.g. ``Immersion'') or landscape (e.g. ``Context'') mode, distributed along a circumference whose radius determines the level of immersion. Triptych configurations consist of separate groups of combined screens, sometimes coupled with additional screens. The robotic arms can also rotate the screens toward viewers standing outside the environment.}~\label{fig:screen-configurations}
   \vspace{-2em}
\end{figure}

\begin{figure}
  \centering
  \includegraphics[width=0.7\columnwidth]{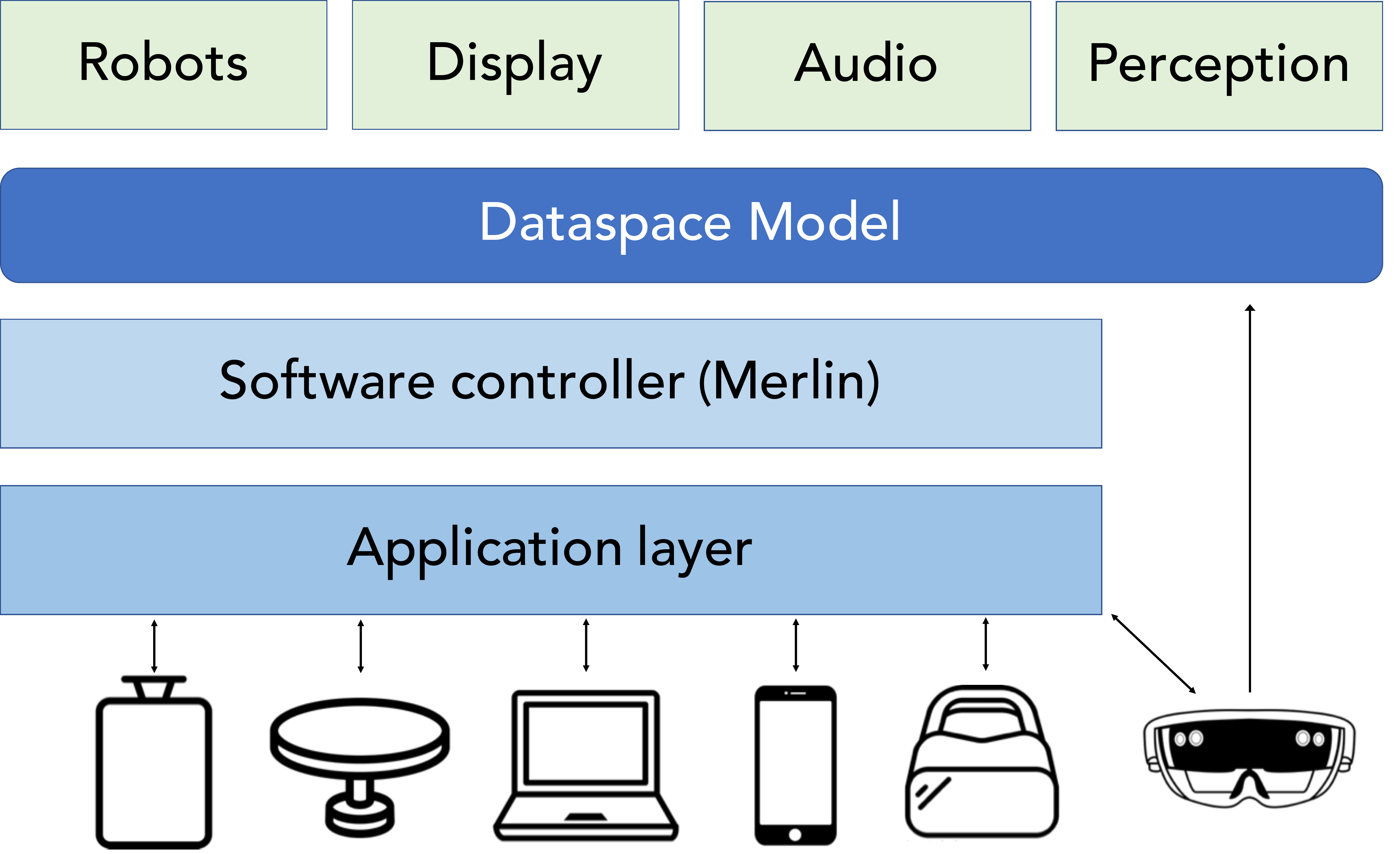}
  \caption{System architecture. Dataspace is composed of four different subsystems whose properties are stored in a spatial representation of the environment. Developers can create web-based applications that access the Dataspace model and connected devices through a software middleware layer called \textit{Merlin}.}~\label{fig:architecture}\\
  \vspace{-2em}
\end{figure}

Dataspace (Fig.~\ref{fig:datacenter}) is a room-sized collaborative environment where multiple researchers can interact naturally with rich and complex data at scale (D1). This environment re-imagines the conference room as a dynamic physical environment that adapts to users, applications, and data.\\
The physical components constituting Dataspace are:
\begin{itemize}[noitemsep,topsep=0pt]
\item 15 OLED 4K resolution displays, which can be reconfigured in space by 15 7-DOF Kuka robotic arms, mounted to the ceiling in a circular pattern with a radius of 2.5m (D2). Due to weight and wiring constraints, the screens do not natively possess touchscreen  capability. However, this functionality is emulated using the torque sensors located in each robot joint (with a sensitivity of $\pm$2cm). While each screen can be moved and rotated independently in space, Figure~\ref{fig:screen-configurations} shows some of the most common configurations (D3). Dataspace possesses both software and hardware safety protocols preventing screens from colliding with each other and with people in the room.

\item A central, smooth-surface ceramic table that can be raised or lowered based on application requirements. Thanks to two HD projectors, blended 2K resolution visual output can be projected onto the table. Despite its surface being passive, table gestures performed by users can be detected through the Dataspace perception system.

\item A central cowling mounted to the ceiling, holding the two projectors mentioned above and eight crossfiring Kinect v2 depth sensors, used by the Dataspace perception system to track objects and people, and identify gestures in the area between the table and the screens.

\item A spatial audio system consisting of a speaker mounted on each robotic arm, plus five speakers and two subwoofers in the cowling. The system also includes four phased-array microphones, which can detect the azimuth of the strongest speech signal and interpret it through IBM Speech-to-text and Conversation services \cite{ibm-cloud}.

\item A set of ten Microsoft Hololens augmented reality headsets, which seamlessly integrate with the other Dataspace components for interaction, audio services and graphical rendering. These AR headsets can be worn by users interested in interacting with 3D information, often displayed atop the central table (D3).

\item An optional, varying set of mobile devices such as laptops, tablets, smartphones and virtual reality headsets (D5), all of which may be used as extensions of Dataspace screens, or as means for users to virtually join the session.

\end{itemize} 

{\setlength{\parindent}{0cm}
Dataspace implements an API-based modular software architecture (Fig.~\ref{fig:architecture}) for handling its subsystems: motion control (coordinating the 15 robotic arms), perception (detecting and tracking people and objects), display (controlling screen content and table projections) and audio (speech interpretation and audio output). Each subsystem runs on a dedicated server machine and communicates with the centralized software controller \textit{Merlin} through a publish/subscribe (MQTT) protocol. We leverage ROS (Robot Operating System) for controlling the robotic arms and IBM Cloud services \cite{ibm-cloud} for speech interpretation, while we developed a custom algorithm for merging the depth information from multiple Kinect devices into a single point cloud.
We chose to adopt web-based rendering for our screens through Electron \cite{electron}, promoting high flexibility for application development (e.g. HTML, WebGL) and support for external devices with browsing capabilities (e.g. laptops and smartphones), so that the same content and interactions can be easily transferred across devices. 
Applications can access the Merlin API through a dedicated Node.js package and be automatically deployed to Dataspace as Docker containers. Alternatively, custom applications can be built by leveraging a RESTful version of the Merlin API.
}

\subsection{Spatial Awareness}

\begin{figure}
\centering
  \includegraphics[width=\columnwidth]{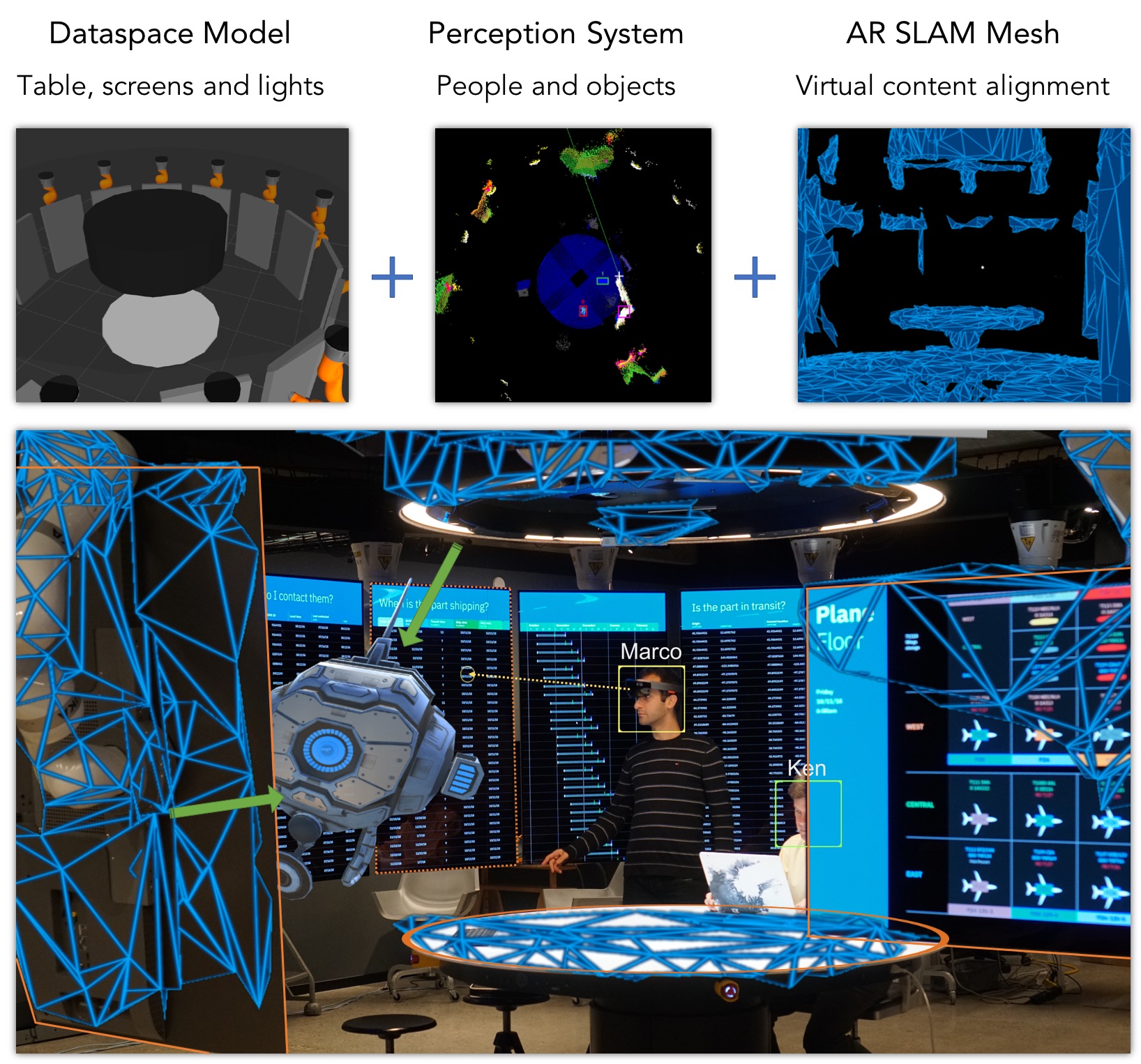}
 \caption{Environmental perception. The 3D model of the environment stored by the robotic arm controller is combined with the point cloud generated by the eight crossfiring Kinect v2 sensors, establishing positions of objects and people with respect to the current screen configuration. This is aligned with the real-time environment mesh built by the AR headsets, enabling correct occlusion rendering and spatial interaction with virtual content.}~\label{fig:spatial}
  \vspace{-2em}
\end{figure}

\begin{figure*}
  \centering
 \includegraphics[width=1\textwidth]{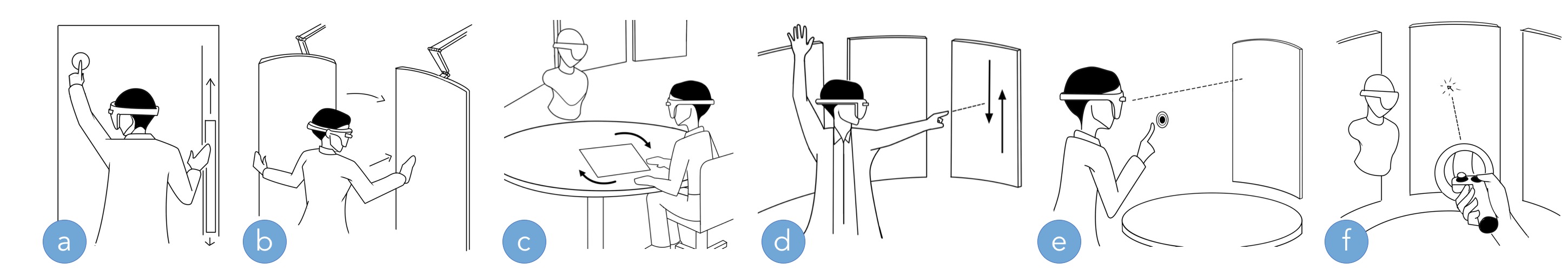}
 \caption{Drawings of the primary interactions possible with  Dataspace. The figure above displays, from left to right, a) touch and b) physical interaction with screens, c) table and d) mid-air gestures, e) AR gaze and gesture integration, and f) VR wand-based extension for remote participants. Despite that the user is shown wearing an AR headset in several of these  interactions, the device is only required in interaction (e). }~\label{fig:interactions}
  \vspace{-2em}
\end{figure*}

In a complex environment mixing people with the dynamic configuration of heterogeneous devices, it is important to maintain a stateful representation of each entity in order to guarantee proper interaction with the content.
Dataspace holds a virtual 3D model of its physical environment (screens, table, robotic arms, and cowling), that is updated by the robotic arm controller every time a new screen configuration is applied.
The point clouds generated by the eight crossfiring Kinect v2 sensors are combined as to produce a single depth representation of the environment, which we apply to the corresponding reference frame in the virtual 3D model. This custom perception system allows us to not only detect and track people and objects in space, but also to determine their relative and absolute position with respect to the table and the screens. For instance, we can easily detect if a person is pointing at a particular screen or a specific location on the table, and then forward this information on to the system's application layer, where it is then transformed into a user interaction. Similarly, knowing that a user is physically close to or far away from specific content can be fundamental in determining which action the system should perform (e.g. bringing a screen or moving the content closer the user, or performing an operation on the nearest screen, especially in response to voice commands).
On top of the environment virtual 3D model and the perception system, we include a third, additional layer represented by the SLAM spatial mapping built at run-time by the AR devices. Whenever a Hololens is powered on, it begins constructing a spatial mesh of the surrounding environment, which is then expanded and updated as the device moves in the space. The headset  attempts to match its version of the environment to a reference (precomputed) mesh anchor. When a match is found, and the devices recognizes its position in the environment, the AR application adjusts its coordinate frame accordingly. By combining information from the virtual 3D model and the perception system, each headset can identify which screen the user is looking at or where people or content are located in the room (Fig.~\ref{fig:spatial}). This is also useful for positioning virtual content inside the workspace (e.g. constraining the position of content to accessible areas) and for proper lighting and occlusion computation (e.g. virtual objects are not rendered if positioned behind other objects). Finally, relying on updates from the virtual 3D model is far more reliable than using built-in Hololens spatial mapping, which takes too much time to recompute its mesh when the environment is physically reconfigured.

\subsection{Interacting with the Environment}

Interaction events registered by each of Dataspace subsystems are collected by the Merlin software middleware and broadcasted to all views and connected devices, which can independently decide how to handle them (D4). 

\bfparhead{Touch and Physical Interaction.}
Thanks to our emulation of touchscreen functionality through the robotic arm torque sensors, users can perform single-touch screen operations such as click, drag, scroll, and zoom (Fig.~\ref{fig:interactions}a), which are then forwarded to the single views as Javascript events. Users can also physically interact with the system by manually moving and rotating screens in space (Fig.~\ref{fig:interactions}b), a function made possible by the robotic arms adapting to force applied by the user (``compliance mode''). This mode of interaction has proven to be effective for applications based on slicing operations (e.g. MRI data, multi-video time analysis).

\bfparhead{Gestures and Movement.}
Thanks to its perception system, Dataspace can identify and respond to gestures performed in mid-air or on the table. Raising both hands in front of a screen will clear that screen's content, while raising one hand is interpreted as a request to speak, 
thus orientating content and lighting towards the requester (Fig.~\ref{fig:interactions}d, D2). Similarly, hands and objects on the table are tracked based on horizontal position and height, allowing for the use of props and for the selection, panning, zooming, rotation and rescaling of views and projected shapes (Fig.~\ref{fig:interactions}c). Table interactions are complemented by a rotational input device (Microsoft Surface dial) whose position is tracked in space. This device is useful for creating radial menus and interactions with content distributed over a circular shape.

\bfparhead{Voice and Sound.}
If a user wants to physically interact with a view that is currently distant from his position, Dataspace can move that view, or even the physical screen, closer to the user. However, in most cases, it is simpler to leverage the environment's microphone array, which, combined with the perception system, enables understanding the source and directionality of speech.
We use the Watson Assistant \cite{watson-assistant} service in combination with speech-to-text and text-to-speech to provide agent-based voice interaction.
While dialogue is application-dependent, we use the ``Merlin'' attention keyword for cross-application vocal commands such as moving robotic arms and/or the content of views. The system's multiple speakers can also generate 3D sounds inside the environment, for instance in proximity to a particular screen or person.

\bfparhead{AR Headsets.}
The AR headset we currently use, Microsoft Hololens, allows us to provide additional interactions, responding to user gaze, hand gestures (airtap, bloom), head motion and voice (we plan on integrating eye tracking in the near future as well). Though they are  mainly used to interact with the 3D AR content visualized at the center of the room, headsets are an integral part of the Dataspace system (D3).
For instance, utilizing audio input through the headset's microphone renders it trivial to identify who is speaking (and their position in space), in addition to providing better audio quality. 
In combination with overlay graphics to provide visual feedback, gaze is used to enable users to select screens and other objects in the environment (Fig.~\ref{fig:interactions}e). For instance, while gazing at a screen and performing a long airtap, the contents of that screen are selected, and can be swapped with the contents of another screen.
Finally, despite this still being an experimental feature, AR headsets can also be used to overlay information atop 2D data already displayed on a screen.
As a separate consideration, we note that the smooth surface of the table and the dark color of the screens do not represent an ideal tracking environment for the Hololens. When movement on the table is highly dynamic, there is a chance that AR content may drift in space. To handle these concerns, we have adopted a design solution that involves  displaying a trackable crown (often used as a radial menu) around the table content, as well as graphical cues on the screens to facilitate visual alignment.

\bfparhead{Integration with External Devices.}
Since content in Dataspace is delivered through web-based technologies, it is relatively straightforward to integrate with devices such as personal laptops, tablets and smartphones (D5). Tasks such as typing and coding are easier to perform through standard devices (e.g. a keyboard), and it is fundamental to provide such devices as a complement to native Dataspace interactions. For instance, if a view on a Dataspace screen requires textual input, the user can decide to provide that input by voice, use a virtual touch keyboard on the screen, or move the view to her personal laptop, type with her keyboard, and then send the view back to the Dataspace screen.
For applications requiring higher frequency input, it is also useful to make gesture-based interactions available in close proximity to the user (and independent of his position in the room), providing a good use case for devices such as smartphones and tablets.

\begin{figure}
\centering
  \includegraphics[width=\columnwidth]{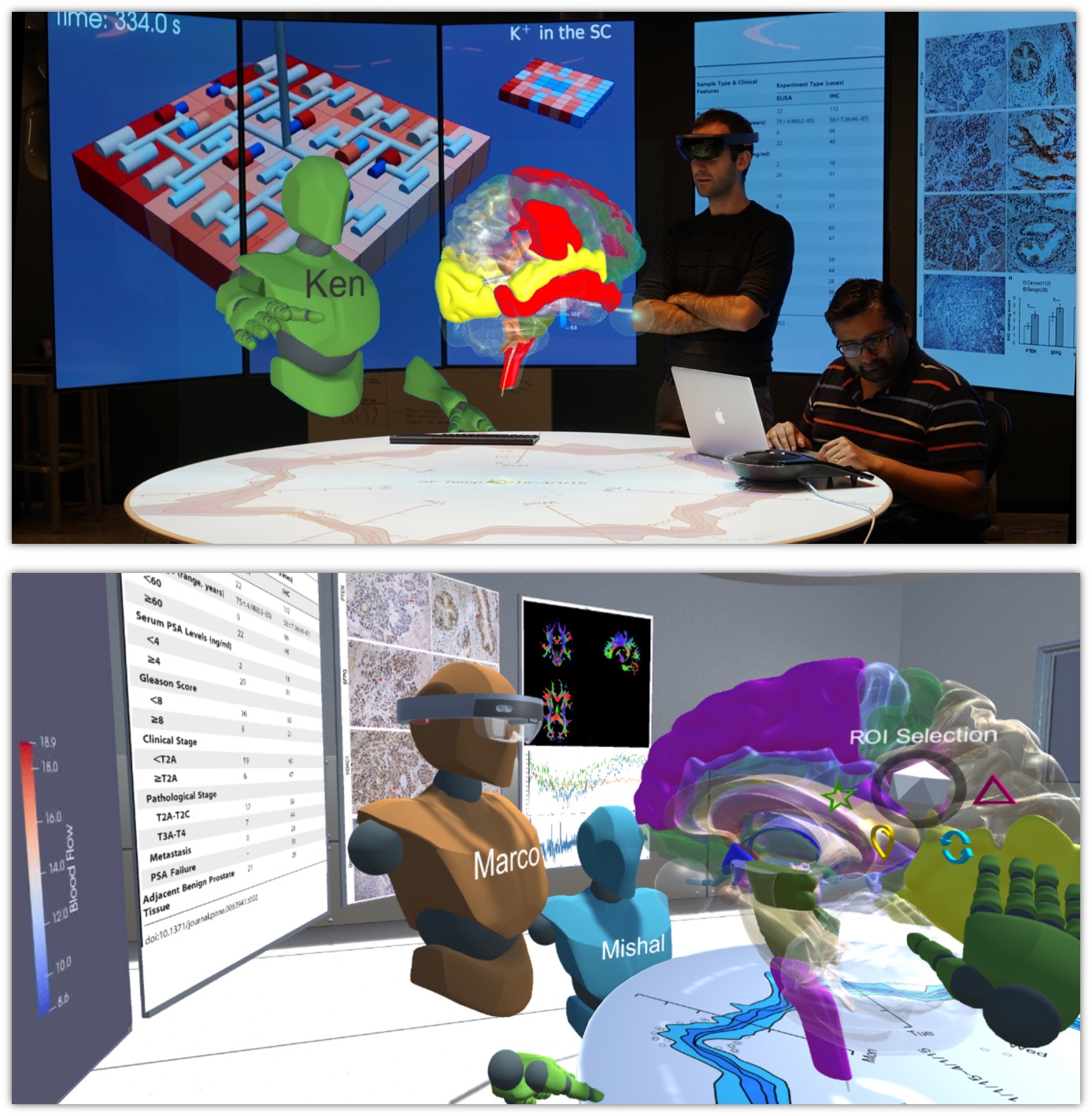}
   \caption{Dataspace virtual reality extension. To enable remote users to participate in data exploration without sacrificing Dataspace's unique capabilities, we built a VR version of the environment that allows scientists to interact in real time with the people physically present in the Dataspace conference room, as well as the content the are viewing. The figures above illustrate how users perceive remote participants through AR (top) and how remote participants interact with the physical room through VR (bottom).}~\label{fig:remote}
  \vspace{-2em}
\end{figure}

\bfparhead{Scalability and Remote Participation.}
Two drawbacks typically associated with immersive environments are price and scalability. In particular, these technologies often require large spaces and a dedicated construction process, and their cost is usually a significant limiting factor in the number of deployed instances.
While in this paper we present Dataspace as a complete system and research environment, its modular design allows for the deployment of any combination of its subsystems. The screens, robotic arms, table, perception system and augmented reality integration can all be independently removed, in which case the system falls back on the interfaces and interaction methods remaining. We have already constructed a second Dataspace system which does not make use of the robotic arms, and can envision versions of the environment with different numbers of screens, based on user need.\\
An interactive virtual reality interactive rendering of Dataspace makes the system available to users in remote locations (D1, D2). To create this rendering, we combine the 3D virtual model updated by the robotic arm controller, the perception system, and AR tracking information to recreate the current physical state of a Dataspace installation. This can be observed in a real time, first person perspective by a  remote user in VR. This VR extension shows the current configuration of the robotic arms, the content displayed on the table and on each screen, and the estimated position of each team member within the space (Fig.~\ref{fig:remote}). The VR user can perform both touch (emulated using the VR hand controllers, as shown in Fig.~\ref{fig:interactions}f) and voice interactions as though they were present in the room. By pulling a virtual screen down with the VR hand controller, for example, both the virtual and the actual screen will move down). Additional  examples of uses for the Dataspace VR extension are described in section \textit{Applications}.

%% file: applications.tex
\section{Applications}

In this section, we illustrate four different Dataspace use cases, all involving hybrid data exploration. For each application, we discuss the advantages of utilizing our environment, comparing methods of interaction with Dataspace to those of existing technologies.

\subsection{Immersive Insights}

\begin{figure}
  \centering
  \includegraphics[width=\columnwidth]{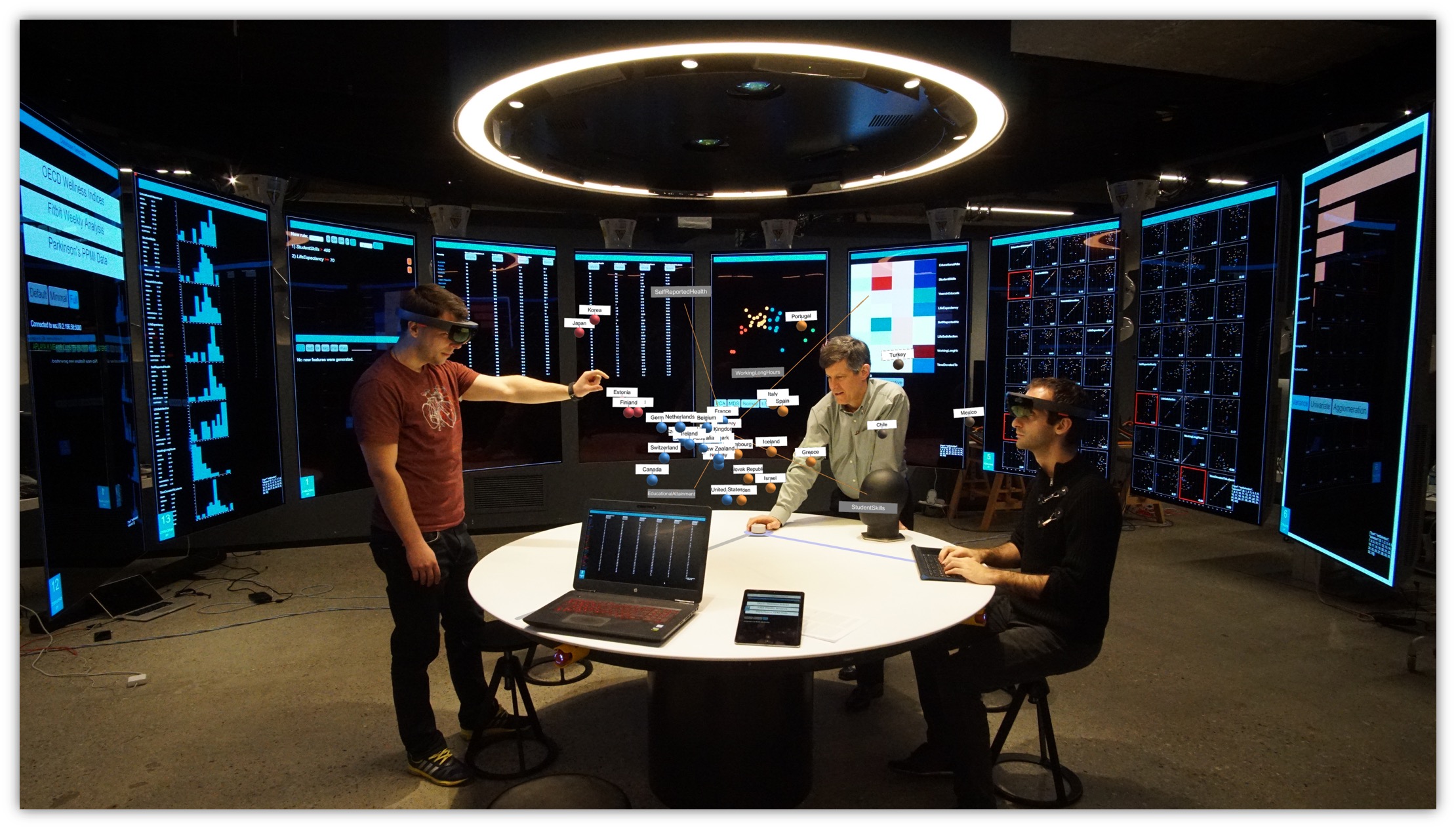}
  \caption{Immersive Insights application. Multidimensional data exploration is a typical use case for immersive environments. In this application we focus on exploratory data analysis, combining views on laptops and Dataspace screens with a central table controller and an AR interactive 3D representation of high-dimensional data.}~\label{fig:immersive-insights}
  \vspace{-2em}
\end{figure}
Exploration of high-dimensional data has always been a popular use case for immersive environments, because they can leverage depth of information to better identify structures and patterns in the data.
Many non-immersive exploratory data analysis (EDA) tools instead  focus on statistical analysis, capitalizing on more interpretable, higher-resolution 2D visualizations.
EDA involves iterating over a large set of possible choices of algorithms and parameters, exploring a vast space of possible solutions which need to be both quantitatively and qualitatively described in order to make sense of the data. In their work on \textit{Clustrophile 2} \cite{clustrophile}, a single-user desktop application for clustering analysis, Cavallo et al. pointed out the need the importance of visually comparing multiple clustering instances characterized by different subsets of the data and algorithms applied, a task limited in effectiveness by screen size.
We conceived of \textit{Immersive Insights} (Fig.~\ref{fig:immersive-insights}) as a natural extension of existing EDA systems - a workspace where multiple data scientists might collaboratively analyze data at scale using a mixture of different devices and forms of interaction, combining the advantages of 3D stereoscopic data exploration with the rigorous statistical analysis typical of 2D visualization tools.
\textit{Immersive Insights} builds upon recent work in exploratory data analysis \cite{clustrophile}, focusing on extracting insights from user provided datasets, and leveraging techniques such as feature selection, clustering and dimensionality reduction.

After uploading or selecting a dataset to analyze, data scientists may choose among several views of the data, displaying them on any of the screens or on users' laptops. Each view is dedicated to a different aspect of the analysis (filtering, projection, clustering, correlation analysis, feature selection and engineering, data distribution, etc.) and can be moved to any screen or resized to occupy multiple screens at a time. For instance, a data table with many columns maybe be extended to use two or three displays, engaging the robotic arms in moving their screens together to form a single, larger display. A user might also decide to move a view from a Dataspace screen to her own laptop, or vice versa.
As Clustrophile 2 introduced the concept of ``clustering views'' \cite{clustrophile} to enable simultaneous analysis of different subsets of the data, Immersive Insights proposes a similar idea, \textit{analysis groups}: groups of screens can be associated with particular data subsets and algorithmic choices, so that data scientists can even work independently on separate tasks, then compare and merge individual results.
Users can perform actions through voice, touch (screens, table) or using a mouse and keyboard (laptop). We utilize IBM Cloud Conversation for parsing and understanding users' requests, and for suggesting specific algorithmic choices or actions based on current data (e.g. if the user says "I would like to apply t-SNE to the data in view number two", the system may reply "The t-SNE algorithm won't perform well on this few data samples, I suggest you to try Isomap instead"). If users are working simultaneously on multiple analysis groups, the system will use directional audio recording to apply actions to the set of screens closest to the person currently speaking.
In Immersive Insights, it is hoped that the view visualized on the table will encourage the team to collectively discuss content and make decisions. This view has  multiple uses: 1) display a radial menu to select screen and load views (or perform operations on them), 2) mirror the perspective of a user currently wearing an AR headset, 3) provide a summary of an analysis and list the current data instances under consideration.
Augmented reality headsets can be used to visualize a dimensionally reduced representation of data as a 3D interactive scatterplot at the center of the table (Fig.~\ref{fig:immersive-insights}). Thanks to the implementation of biplot axes, prolines and forward and backward projection techniques \cite{cavallo2018visual}, the AR scatterplot is fundamental to interactively understanding the importance of data features and clustering assignments in the dimensionally reduced representation. The AR headset further provides visual feedback and spatial assistance while interfacing with the screens (e.g. for extending views or moving them from one screen to another).

\begin{figure}
 \centering
 \includegraphics[width=\columnwidth]{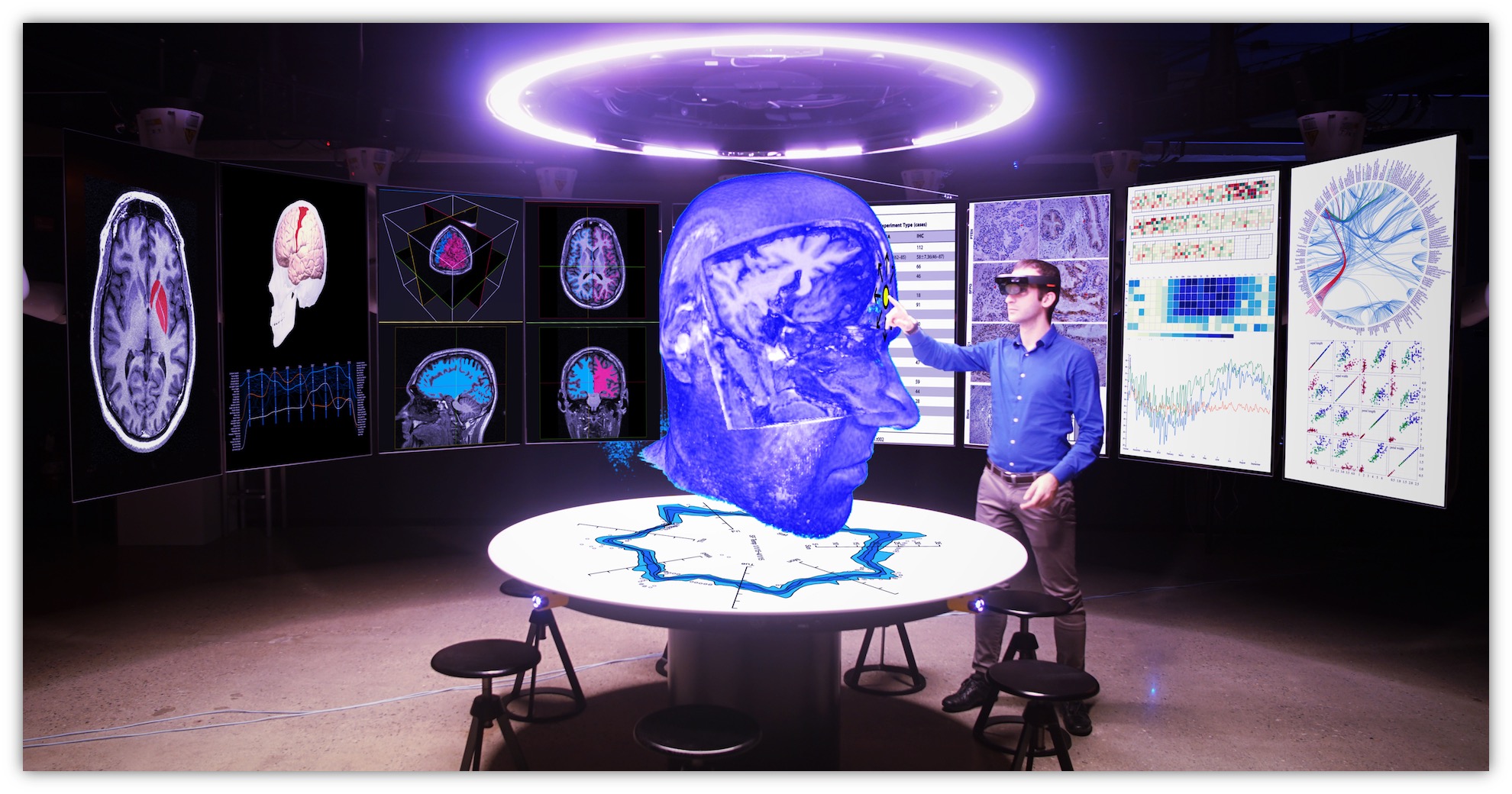}
 \caption{The Holobrain project. The Dataspace acts as a control room receiving requests from different emergency rooms (ERs), which can send real time patient data for analysis and verification. A set of experts can examine from the Dataspace screens the incoming data streams and additional statistics or model outputs, and decide which ones to combine in the central AR brain visualization. The medical experts can then send information back to the ER with feedback on what procedures to take on the patient. }~\label{fig:holobrain}
 \vspace{-2em}
\end{figure}

\begin{figure}
 \centering
 \includegraphics[width=\columnwidth]{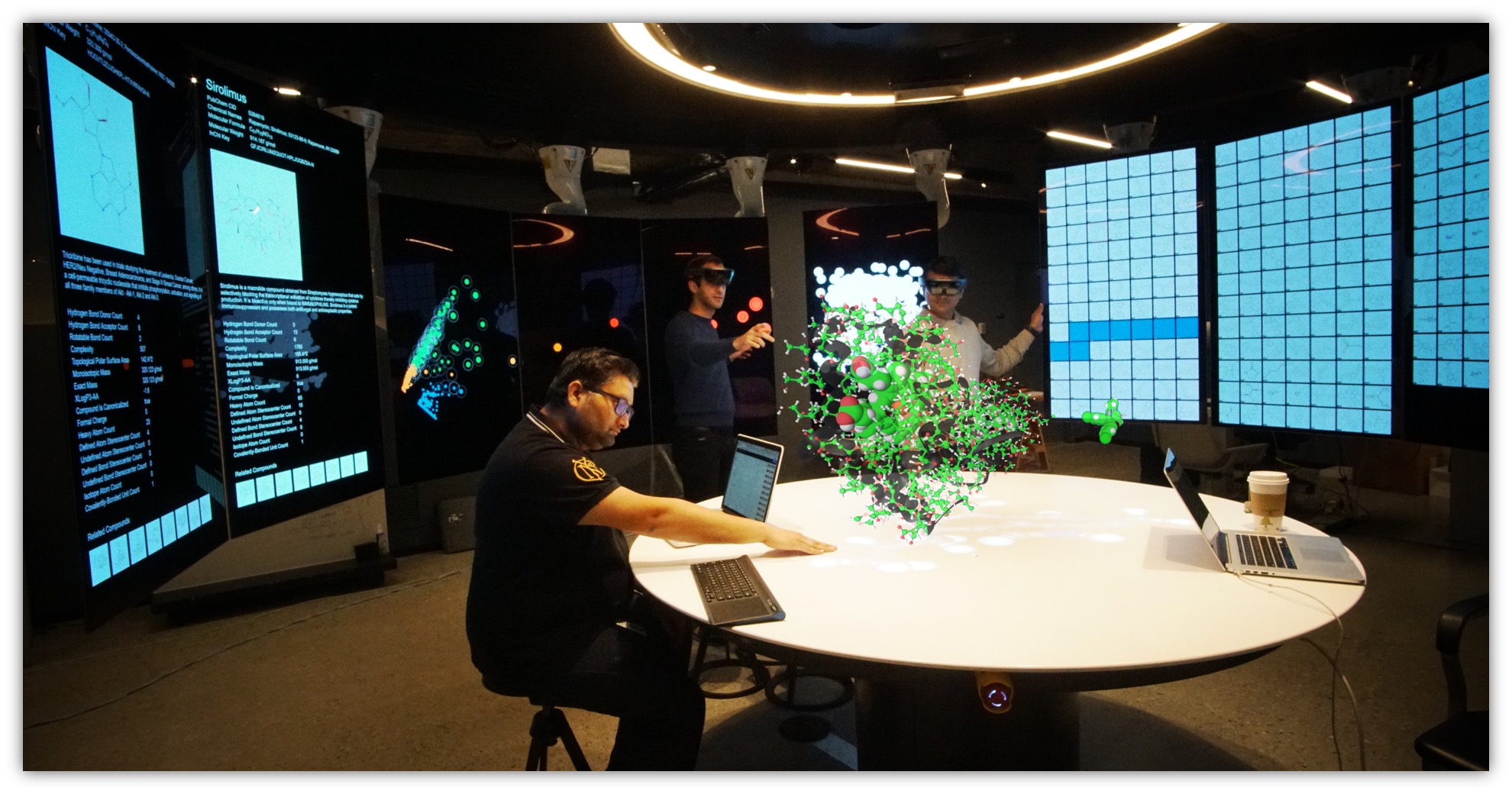}
 \caption{Dataspace Drug Discovery tool. Used by molecular researchers and data scientists, the Drug Discovery tool enables interactive exploration of the various molecules that bind to a desired protein target. In this example, a frontal triptych displays a scatterplot including all molecules of interest, while additional screens can be added on request to provide information on single molecules or summary statistics. Once the molecules of interest have been identified, their 3D structure becomes available to view at the center of the table through AR, enabling interactive hypothesis testing on bindings and the creation of new molecules.}~\label{fig:drug-discovery}
 \vspace{-2em}
\end{figure}

\subsection{Data Center Management}
Another common use case for immersive environments is the analysis of corporate data associated with flight control, warehouse management, internet transactions, or data center monitoring-applications in which temporal events and risk analysis are often related to some form of spatial information.
For example, the monitoring and management of data center facilities is generally characterized by large amounts of data associated with the performance, power consumption and service status of server machines and the software processes they run. In particular, it is important to determine which pieces of hardware are experiencing failures and may need to be replaced, but also to quickly identify which software processes may not be functioning as they should, in order to isolate and understand the root cause of an issue. Geospatial information, such as the physical location of facilities and servers, is generally used to further estimate the causes and impact of each failure. Interactively analyzing data associated with billions of processes running on millions of virtual machines hosted by thousands of physical servers can be computationally and visually difficult on normal laptops.
The Data Center Dataspace application (Fig.~\ref{fig:datacenter}) tries to accommodate these requirements, covering tasks that range from network activity monitoring, fault detection, tracking the status of service processes, and server replacement and relocation.
While presenting 2D information is sufficient to provide summary insights on a data center, spatial information is often  lost in this context. For this reason, we leverage the 3D capabilities of Dataspace to help users quickly identify, on an interactive 3D globe, the geographical location of servers and the countries affected by service interruptions. Similarly, we provide a more granular 3D view of the data center's physical configuration, showing the positions of servers hosting a particular process and visualizing sensor information such as temperature and ventilation, which can be useful in understanding the root cause of certain hardware faults.

\subsection{Brain Data Visualization}

The Holobrain project \cite{holobrain} was originally born as an AR/VR application for visualizing regions of interest in brain atlases. These regions often encompass volumetric parts of the human brain identified as relevant to a particular disease by mathematical simulations or machine learning processes. For instance, Holobrain was originally used to spatially visualize the 13 regions of activation identified in subjects with Schizophrenia \cite{gheiratmand2017learning}.
Increased demand for a way to visualize several types of brain-related sensor data, the biological properties of brains, and machine learning outputs, made the original application visually cluttered and not suited for visualizing certain data streams at a desirable resolution.
With this in mind, we designed a Dataspace version of the Holobrain project as a control room (Fig.~\ref{fig:holobrain}) wherein a group of medical experts might inspect and analyze patient data. In our use case, data is transmitted by clinicians located in remote emergency rooms where they must decide what treatment to apply to patients. Incoming data can consist of brain scans, live sensor recordings and other biological information (e.g. volumetric temperature, blood pressure in brain regions). While this information is displayed on the Dataspace screens, mathematical models can be run in parallel to generate insights on the patient and suggest possible courses of action. Medical experts can decide which incoming data streams and computed information might be relevant, and choose to display their selections in a central AR brain visualization, mapping relevant data to a brain atlas (or the patient's brain, when available). Here, the user can interactively open the virtual brain and slice it in the direction of interest, attempting to reveal important regions, and deciding, based on brain structure and overlaid information, what is the best course of action for the patient. Finally, this feedback may be transmitted to the clinician in the emergency room.
Despite that it is not viable to host a Dataspace installation in each hospital, it is instead perfectly reasonable to include a small VR headset in an emergency room, so that the clinician may have a direct connection with the remote Dataspace control room, thereby  conducting an interactive consulting session with remote medical experts (Fig. \ref{fig:remote}).

\input{comparison}
\subsection{Drug Discovery}

Another typical use case of immersive environments is represented by the observation and analysis of biological phenomena, in particular molecular structures. While CAVE  has proved to be reasonably effective in 3D molecular visualization \cite{febretti2013cave2}, Dataspace extends this use case to simultaneous 2D and 3D analysis of both biomedical datasets and chemical structures.
In drug discovery, researchers identify which small molecules (compounds) can be bound to a protein target associated with a particular symptom or disease, hoping to mitigate or cure negative effects. This process generally involves several steps, including: 1) identify which genes and protein targets are known to be related to a particular disease, 2) explore a vast, multidimensional set of molecules that may bind to the protein target, 3) filter that set based on molecular parameters, existing patents and the side effects caused by each molecule, 4) identify reasonable trade-offs and try to generate new molecules, leveraging the 3D structural properties of the molecules of interest. While the last step usually requires that researchers visualize the position of single atoms in 3D space, the remaining steps of the analysis require a mixture of several information-heavy 2D visualizations and images, often using data extracted from large online medical datasets that are difficult to fit on an ordinary laptop.
After gathering requirements for a group of molecular modeling scientists, we began developing a drug discovery application (Fig.~\ref{fig:drug-discovery}) that leverages the large, multi-screen surface provided by Dataspace, while maintaining the possibility of switching to a dedicated AR view of the 3D structural content of molecules of interest. This way, researchers can, in the same application, analyze both large amounts of textual/numerical information as well as molecular 3D properties, maintaining good resolution and readability for both.

%% file: comparison.tex
\bgroup
\def\arraystretch{1.2}
\begin{table*}[!htbp]
\resizebox{\textwidth}{!}{
\centering
\begin{tabular}{lccccc}
\toprule
                     & \textbf{CAVE}                   & \textbf{CAVE2}                   & \textbf{Reality Deck}                      & \textbf{CEL}                & \textbf{Dataspace}                                \\
\textbf{Year}        & 1992                            & 2012                             & 2015                                       & 2016                        & 2018                                              \\
\textbf{Resolution}  & 2.6 Mpx                         & 36 Mpx                           & $\sim$1.5 Gpx                              & 33.2 (+16) Mpx                    & $\sim$129 Mpx                                     \\
\textbf{Dimensions}  & 10 x 10 x 10ft                & 24ft diameter, 8ft tall          & 33
× 19 × 11ft                    &   13.36ft wide, 7.5ft tall       & 16ft diameter, 9ft tall                           \\
\textbf{Graphics}    & 4 projectors, 6 walls           & 72 LCDs                          & 416 LCDs                                   & 16 (+5) LCDs                     & 15 4K LCDs, 2 projectors, AR headsets             \\
\textbf{Rendering}   & Stereo 3D                       & Hybrid 2D/3D                     & 2D with depth cues                         & 2D                          & Hybrid 2D/3D, personalized                        \\
\textbf{Processing}  & 4x MIPS R4000, SGI Crimson VGXT & 36x Xeon E5-2690, Nvidia GTX 680 & 18x Xeon E5645 nodes, 4 FirePro V9800 each &           4x Intel Xeon 2667, NVIDIA K5000                  &                                                  
8x Xeon E5-2699, 11x NVIDIA Quadro M6000 \\
\textbf{Interaction} & 3D wand                         & 3D wand                          &                        N/A                    & 3D wand, voice              & Touch, 3D gestures, voice, dial, external devices \\
\textbf{Tracking}    & Markers                         & Markers (14-camera Optitrack)    & Markers (24-camera optitrack)              & Markerless (2 VIVE cameras) & Markerless (8 Kinect v2)                          \\
\textbf{Audio}       & (1 for each corner)             & 20.2 surround                    & 24.4 surround                              &           8 speakers, 20 microphones                  & 20.2 surround, 4 microphones                                    \\
\textbf{Props}       & None                            & (Desk, chairs)                   & None                                       & None                        & Robots, table, chairs, mobile devices             \\
\textbf{Reconfig.}   & No                              & No                               & No                                         & No                          & Yes                                               \\
\textbf{Agents}      & No                              & No                               & No                                         & Celia                       & Merlin (Watson)                                   \\
\textbf{Remoting}    & No                              & 2D video feed, SAGE2             & No                                         & No                          & 2D video feed, AR avatar, VR extension            \\
\textbf{Cost (USD)}  & \$2M                            & \$926K                           & \$950K                                     &             ~\$750K                & $\sim$\$1M for all modules

\\
\bottomrule
\end{tabular}
}
\vspace{0.1em}
\caption{Table comparing Dataspace to previous immersive environments. Dataspace mostly differentiates itself in its hybrid rendering, reconfigurable design, device integration, modularity and extensibility. We note that costs refer to the year in which each environment was first built, and are expressed in today's USD value.}
\label{table:comparison}
\vspace{-2em}
\end{table*}
\egroup

%% file: discussion.tex
\section{Discussion}

\subsection{\xcancel{Immersive} Hybrid Analytics}

The design of collaborative, immersive environments for data exploration has historically been based on balancing the complex technological trade-offs among hardware complexity, image quality, resolution, field of view, depth rendering, visual acuity, perception issues (e.g. ghosting), and cost \cite{manjrekar2014cave}. Originally, CAVE was born as a small environment for 3D scientific exploration, applied to domains such as biology, fluid dynamics, architecture and geospatial data. CAVE2 and Reality Deck introduced multi-user data exploration, outlining a space large enough to promote user movement. In the former case, simultaneous rendering of 2D and 3D scientific content was introduced, whereas the latter has focused on ultra high resolution immersive graphics, foregoing the use of stereoscopic 3D and adopting depth clues (e.g. motion parallax) instead. A technology comparison table is presented in Table~\ref{table:comparison} for reference.

With its capability to simultaneously render 2D and 3D content, Dataspace represents a second implementation of the concept of a \textit{hybrid reality environment}, after CAVE2. At the same time, the pixel density of Dataspace is comparable to that of Reality Deck, but without having Reality Deck's almost 360\textdegree horizontal FOV coverage. This technological difference makes Dataspace less applicable to immersive scientific exploration (i.e. our environment cannot completely surround the user with visuals according to the original criteria proposed by Cruz et al. \cite{cruz1992cave}). 
However, Dataspace provides ``outside-in'' 3D stereoscopic rendering combined with high resolution 2D visual analytic capabilities, enabling what we call \textit{hybrid analytics}.
Whereas previous environments have mostly dealt with visual exploration of scientific data, the unique characteristics of Dataspace make it a good candidate for flexible focus-and-context analysis of multiple types of information at the same time --- as demonstrated by the applications presented in the previous section. 
In particular, Dataspace bridges complementary visualization environments and allows users to seamlessly switch between them, providing a wider support for the different perceptual and interaction tasks that characterize visual information analysis \cite{bach2018hologram}.

\subsection{Towards the Meeting Room of the Future}

While in this work we presented Dataspace as a research environment, it is our goal to continue exploring how its components can be better combined.
We briefly discussed how the possibility of reconfiguring screens in space applies differently across application scenarios. However, we would like to even further explore how screens can be dynamically reconfigured in space within the same application and how physical interaction (i.e. moving screens by hand) can be better exploited.
Similarly, we hope to further explore the concept of egalitarian access to the data, a key design factor differentiating Dataspace from other immersive environments. In particular, how will gestures (e.g. raising your hand) and interfaces (e.g. the puck) be used in taking control, sharing and redirecting content within a group of people?
In terms of collaboration, the possibility of decoupling visualizations and interactions and separately providing them to different users or groups of users represents an interesting challenge, requiring further study as to how people will interact with each other and combine their individual results while using the same application, in the same physical space.
Finally, it is our mission to continue exploring how AR and VR technologies can come into play in these contexts, eventually becoming integral to collaborative environments.

%% file: conclusion.tex
\section{Conclusion}

In this paper we introduced a new hybrid environment called Dataspace, which aims at exploring new types of interaction in collaborative environments. In particular, Dataspace focuses on a seamless integration of different types of technology, offering a hybrid approach to the delivery of immersive analytics.
 Our discussion focused on the integration of physical workspaces with AR and VR technologies, which proved to be a fundamental extension for handling specific types of data and mitigating system scalability issues.
We also demonstrated, through four real-world applications, how Dataspace can be used in very different domains and adapted to different user requirements, and we examined  the advantages and trade-offs of the system compared with  existing technologies.
We believe this research will be helpful in developing better collaborative workspaces.